\documentstyle[11pt,titlepage]{article}
\renewcommand{\arraystretch}{1.4}
\input amssym.def
\input amssym.tex

\begin{document}

\pagestyle{plain}
\pagenumbering{arabic}
\def\thesection{\Roman{section}.}
\def\thesubsection{\Roman{section}-\Alph{subsection}.}
\def\theequation{\Roman{section}.\arabic{equation}}
\def\<{\begin{equation}}
\def\>{\end{equation}}
\def\intd{\displaystyle\int}
\def\rf#1{(\ref{#1})}
\def\e#1{\mbox{e}^{#1}}
\def\sgn{\mathop{\rm sgn}\nolimits}
\font\Bbb=msbm10
\def\R{\mbox{\Bbb R}}
\def\C{\mbox{\Bbb C}}
\def\P{{\cal P}}
\def\Ph{\widehat{\P}}
\def\Pb{\overline{\P}}
\def\kh{\widehat{\chi}}
\def\hb{\overline{h}}
\def\hbt{\widetilde{\hb}}
\def\Kb{\overline{\!K}}
\def\Kbt{\widetilde{\Kb}}
\def\Omegab{\overline{\Omega}}
\def\qp{\overrightarrow{q},\overrightarrow{p}}
\def\rhoh{\widehat{\rho}}
\def\build#1#2{\mathrel{\mathop{\kern 0pt#1}\limits_{#2}}}

\renewcommand{\thefootnote}{\fnsymbol{footnote}}
\title{\vspace*{-2cm}
\hfill TIFR/TH/02-15 \\
\hfill PM/02-15 \\
\vfill
\LARGE\bf Bell inequalities in four dimensional phase space and the
three marginal theorem\thanks{Work supported by the Indo-French
Centre for the Promotion of Advanced Research, Project Nb 1501-02.}.}
\author{{\Large G. Auberson}\thanks{e-mail: auberson@lpm.univ-monpt2.fr} \\
	\sl Laboratoire de Physique Math\'ematique, UMR 5825-CNRS, \\
	\sl Universit\'e Montpellier II, \\
	\sl F-34095 Montpellier, Cedex 05, FRANCE.
\and
	{\Large G. Mahoux}\thanks{e-mail: mahoux@spht.saclay.cea.fr} \\
	\sl Service de Physique Th\'eorique, \\
	\sl Centre d'\'Etudes Nucl\'eaires de Saclay, \\
	\sl F-91191 Gif-sur-Yvette Cedex, FRANCE.
\and
	{\Large S.M. Roy}\thanks{e-mail: shasanka@theory.tifr.res.in} and
	{\Large Virendra Singh}\thanks{e-mail: vsingh@theory.tifr.res.in} \\
	\sl Department of Theoretical Physics, \\
	\sl Tata Institute of Fundamental Research, \\
	\sl Homi Bhabha Road, Mumbai 400 005, INDIA.
}
\date{May 28, 2002.}
\maketitle
\renewcommand{\thefootnote}{\arabic{footnote}}

\begin{abstract}
We address the classical and quantum marginal problems,
namely the question of simultaneous realizability through a common
probability density in phase space of a given set of compatible
probability distributions. We consider only distributions authorized
by quantum mechanics, i.e. those corresponding to complete commuting sets
of observables. For four-dimensional phase space with position
variables $\vec q$ and momentum variables $\vec p$, we establish the
two following points: i) given {\sl four} compatible probabilities for
($q_1,q_2$), ($q_1,p_2$), ($p_1,q_2$) and ($p_1,p_2$), there does not
always exist a positive phase space density $\rho(\vec q, \vec p)$
reproducing them as marginals; this settles a long standing
conjecture; it is achieved by first deriving Bell-like inequalities in
phase space which have their own theoretical and experimental
interest. ii) given instead at most {\sl three} compatible probabilities,
there always exist an associated phase space density $\rho(\vec q,
\vec p)$; the solution is not unique and its general form is worked
out. These two points constitute our ``three marginal theorem''.

\vspace{1cm}
\noindent PACS : 03.65.Ta, 03.67.-a
\end{abstract}

\section{Introduction}
\setcounter{equation}{0}

In classical mechanics position and momentum can be simultaneously specified.
Hence phase space density has a well defined meaning in classical statistical
mechanics.  In quantum theory the probability density for observing
eigenvalues of a complete commuting set (CCS) of observables is specific to
the experimental context for measuring that CCS.  Joint probabilities for
different CCS which contain mutually noncommuting operators are not defined.
For example for a 2-dimensional configuration space, with $\vec q, \vec p$
denoting position and momentum, probability densities of
anyone of the four CCS $(q_1,q_2)$, $(q_1,p_2)$, $(p_1,q_2)$ or $(p_1,p_2)$
are defined, but not their joint probabilities.  The question one may raise
is : can one define such joint probabilities, e.g. a phase space
probability density $\rho(\vec q,\vec p)$ such that all its
marginals\footnote{ In agreement with common terminology, by
``marginal'' of a distribution over several variables, we denote
integrals of the distribution over a subset of its variables.} coincide
with the quantum mechanical probabilities for the different
individual CCS? This general question was first raised by Martin
and Roy \cite{MartinRoy1}.

The Martin-Roy contextuality theorem demonstrates the impossibility of
realizing quantum probability densities of all possible choices of the
CCS of observables as marginals of one positive definite phase space
density. For example, consider a two dimensional configuration space.
Let coordinates $q_{1\alpha}$, $q_{2\alpha}$ be obtained from
$q_1,q_2$ by a rotation of arbitrary angle $\alpha$, and momenta
$p_{1\alpha}$, $p_{2\alpha}$ be related similarly to $p_1,p_2$
\begin{eqnarray}
  \left(\matrix{q_{1\alpha} \cr q_{2\alpha}}\right) = V\left(\matrix{q_1
  \cr q_2}\right), \ \left(\matrix{p_{1\alpha} \cr p_{2\alpha}}\right) =
  V\left(\matrix{p_1 \cr p_2}\right),
  \label{I.1}
\end{eqnarray}
where
\begin{eqnarray}
  V = \left(\matrix{\cos\alpha & \sin\alpha \cr -\sin\alpha &
  \cos\alpha}\right). 
  \label{I.2}
\end{eqnarray}
Does there exist for every quantum state (with density operator
$\rhoh$) a positive definite phase space density $\rho (\vec q, \vec
p)$ such that its marginals agree with the corresponding quantum
probabilities, i.e.,
\begin{eqnarray}
  \int dp_{1\alpha} dq_{2\alpha}\ \rho (\vec q,\vec p) = \langle
  q_{1\alpha}, p_{2\alpha} |\,\rhoh\,| q_{1\alpha},p_{2\alpha}\rangle
  \label{I.3}
\end{eqnarray}
for all $\alpha$ ranging from 0 to $2\pi$? They answered this question in the
negative by finding a state $\rhoh$ for which eqs.~\rf{I.3} for all
$\alpha$ are inconsistent with positivity of $\rho$.  Since different
$\alpha$ correspond to different experimental contexts, the Martin-Roy
theorem is a new Gleason-Kochen-Specker type contextuality
theorem \cite{GKS2}. The positivity of the phase space density
$\rho(\vec q,\vec p)$ is absolutely 
crucial for this theorem; otherwise the Wigner distribution
function \cite{Wigner3} would be a solution of \rf{I.3}.

Equations \rf{I.3} constitute conditions on an infinite set of marginals
of $\rho(\vec q,\vec p)$ (corresponding to the continuously infinite
choices for $\alpha$) to agree with corresponding quantum probability
densities.  Their inconsistency still leaves open the question of
consistency of a finite number of such marginal conditions.  

Indeed, the consistency of two marginal conditions where the marginals
involve only nonintersecting sets of variables has been known for some time.  
Cohen and Zaparovanny \cite{CZ4} constructed the most general positive
$\rho(\vec q,\vec p)$ obeying
\[
  \int d\vec p\ \rho(\vec q,\vec p) = \langle \vec q|\rhoh|\vec q\rangle, \
  \int d\vec q\ \rho(\vec q,\vec p) = \langle \vec p|\rhoh|\vec p\rangle.
\]
Their solutions generalize the obvious simple uncorrelated solution for pure
states $\psi$,
\[
  \rho(\vec q,\vec p) = |\psi(\vec q)|^2\ |\tilde\psi(\vec p)|^2,
\]
where tilda denotes Fourier transform.  Based on generalized phase space
densities exhibiting position momentum correlations, Roy and Singh \cite{RS5}
constructed a causal quantum mechanics reproducing quantum position and
momentum probability densities, thus improving on De Broglie-Bohm
mechanics \cite{DBB6} which only reproduced the quantum position probability
densities.  Later, going much further than the nonintersecting marginals of
Cohen et al. \cite{CZ4}, Roy and Singh \cite{RS7} constructed a causal quantum
mechanics based on a positive $\rho(\vec q,\vec p)$ whose marginals reproduce
the quantum probability densities of a chain of $N+1$ different CCS, e.g.
\[
  (Q_1,Q_2,\cdots,Q_N), \ (P_1,Q_2,\cdots,Q_N), \
  (P_1,P_2,Q_3,\cdots,Q_N), \cdots (P_1,P_2,\cdots,P_N).
\]
Here $N$ is the dimension of the configuration space, and each CCS in the
chain is obtained from the preceding one by replacing one of the position
operators $Q_i$ by the conjugate momentum operator $P_i$.

Roy and Singh proposed the following definition: a {\bf Maximally
Realistic Causal Quantum Mechanics} is a causal mechanics which
simultaneously reproduces the quantum probability densities of the
maximum number of different (mutually noncommuting) CCS of observables
as marginals of the same positive definite phase space density.  They
also conjectured that for $N$ dimensional configuration space this maximum number
is $N+1$.

A proof of this long standing conjecture is important for quantum mechanics
where it quantifies the extent of simultaneous realizability of non commuting
CCS.

In this paper, we restrict ourselves to the case of $N=2$ degrees of
freedom. The general case ($N>2$) will be dealt with in a forthcoming
paper. In Section II below, we first state the {\bf classical and
quantum marginal problems} and second, show that, given four classical
compatible two-variable probability distributions, there does not
always exist a positive phase space distribution reproducing them as
marginals. In Section III, we develop a new tool, ``the phase space
Bell inequalities'', which are the phase space analogues of the
standard Bell inequalities \cite{JSB8} for a system of two spin-half
particles. We use them in Section IV to prove the conjecture for
four-dimensional phase space ($N=2$), namely the impossibility of
simultaneous realization of quantum probabilities of more than three
CCS as marginals. In Section V, we explicitly construct the most general
phase space distribution which reproduces probabilities of three CSS
as marginals. These results, the {\bf three marginal theorem}, are
relevant for the construction of maximally realistic quantum mechanics.

As our results are essentially new theorems for multidimensional
Fourier transforms, they are also expected to be useful for classical
signal and image processing~\cite{Cohen}. The theorems of the present
paper and their generalizations to arbitrary $N$ \cite{AMRS2N}
considerably advance previous results in the field, which have only
dealt with nonintersecting sets of marginals (e.g. time and
frequency). A summary of the results of this paper without detailed
proofs is being reported separately \cite{AMRSletter}.

\section{Four marginal problem}
\setcounter{equation}{0}

Let us consider a physical system with 2-dimensional configuration space. Let
$(q_1,p_1)$ and $(q_2,p_2)$ be a set of canonical variables in the
corresponding phase space. We look for a (normalized) probability
distribution $\rho(q_1,q_2,p_1,p_2)$ such that
\< \rho(q_1,q_2,p_1,p_2) \geq 0\,,
\>
\begin{eqnarray}
  \int dp_1dp_2\,\rho(q_1,q_2,p_1,p_2) & = & R(q_1,q_2)\,, \label{II.2.R} \\
  \int dp_1dq_2\,\rho(q_1,q_2,p_1,p_2) & = & S(q_1,p_2)\,, \label{II.2.S} \\
  \int dq_1dp_2\,\rho(q_1,q_2,p_1,p_2) & = & T(p_1,q_2)\,, \label{II.2.T} \\
  \int dq_1dq_2\,\rho(q_1,q_2,p_1,p_2) & = & U(p_1,p_2)\,, \label{II.2.U}
\end{eqnarray}
where the four marginals $R(q_1,q_2)$, $S(q_1,p_2)$, $T(p_1,q_2)$, and
$U(p_1,p_2)$, the respective joint probabilities, are given. For consistency
we must have
\< R,S,T,U\geq0\,, \label{II.3} \>
and
\<\begin{array}{rcl}
  \intd dq_2\,R(q_1,q_2)&=&\intd dp_2\,S(q_1,p_2)\,,\\
  \intd dq_1\,R(q_1,q_2)&=&\intd dp_1\,T(p_1,q_2)\,,\\
  \intd dq_1\,S(q_1,p_2)&=&\intd dp_1\,U(p_1,p_2)\,,\\
  \intd dq_2\,T(p_1,q_2)&=&\intd dp_2\,U(p_1,p_2)\,.
\end{array} \label{II.4} \>

	We shall refer to the problem {\sl Given four distributions $R$, $S$,
$T$ and $U$, satisfying the consistency conditions, does there always exist a
positive $\rho(q_1,q_2,p_1,p_2)$ with these distributions as marginals?} as
the {\sl Classical four marginal problem}.

	When the system is quantum mechanical and is described by a state
vector $|\Psi\rangle$, each of the four marginals involves a pair of
compatible observables and we have
\<\begin{array}{rcl}
  R(q_1,q_2) & = & |\langle q_1,q_2|\rhoh|q_1,q_2\rangle|^2\,, \\
  S(q_1,p_2) & = & |\langle q_1,p_2|\rhoh|q_1,p_2\rangle|^2\,, \\
  T(p_1,q_2) & = & |\langle p_1,q_2|\rhoh|p_1,q_2\rangle|^2\,, \\
  U(p_1,p_2) & = & |\langle p_1,p_2|\rhoh|p_1,p_2\rangle|^2\,.
\end{array} \label{II.5} \>
In this case, the above consistency conditions are automatically satisfied.
We then refer to the problem as the {\sl Quantum four marginal problem}. A
positive answer to it for all states $\rhoh$ would mean that a
realistic interpretation of the quantum results is possible (to the
extent that only measurements connected to the four marginals are involved).

We shall see that the answer to both problems is negative.

	Let us first show that the classical four marginal problem does not
always admit a solution. To this end, consider the following set of marginals
\begin{eqnarray}
  R(q_1,q_2) & = & {1\over2}\left[\delta(q_1-a_1)\delta(q_2-a_2)+
    \delta(q_1-a'_1)\delta(q_2-a'_2)\right]\,, \label{II.6.R} \\
  S(q_1,p_2) & = & {1\over2}\left[\delta(q_1-a_1)\delta(p_2-b_2)+
    \delta(q_1-a'_1)\delta(p_2-b'_2)\right]\,, \label{II.6.S} \\
  T(p_1,q_2) & = & {1\over2}\left[\delta(p_1-b_1)\delta(q_2-a_2)+
    \delta(p_1-b'_1)\delta(q_2-a'_2)\right]\,, \label{II.6.T} \\
  U(p_1,p_2) & = & {1\over2}\left[\delta(p_1-b_1)\delta(p_2-b'_2)+
    \delta(p_1-b'_1)\delta(p_2-b_2)\right]\,, \label{II.6.U}
\end{eqnarray}
which obviously satisfy the consistency conditions \rf{II.3} and \rf{II.4}.
They possess two essential features. First, their non factorized form.
Second, in view of the expressions of $R$, $S$ and $T$, the positions of the
factors $\delta(p_2-b_2)$ and $\delta(p_2-b'_2)$ in the expression of $U$
are not the ``natural ones''.

	Eq.\rf{II.6.R} means that the support of the distribution $R$ in the
plane $(q_1,q_2)$ consists in the two points $(a_1,a_2)$ and $(a'_1,a'_2)$.
As a consequence, any positive $\rho$ satisfying \rf{II.2.R} should have a
support the projection of which on the plane $(q_1,q_2)$ would also consist
in those two points. That is
\< \rho = \delta(q_1-a_1)\delta(q_2-a_2)\alpha(p_1,p_2)+ 
   \delta(q_1-a'_1)\delta(q_2-a'_2)\alpha'(p_1,p_2)\,, \label{II.7.R}
\>
where $\alpha$ and $\alpha'$ are some positive distributions. Similarly, from
eqs.\rf{II.2.S} to \rf{II.2.U}
\begin{eqnarray}
   \rho & = & \delta(q_1-a_1)\delta(p_2-b_2)\beta(p_1,q_2)+
   \delta(q_1-a'_1)\delta(p_2-b'_2)\beta'(p_1,q_2)\,, \label{II.7.S} \\
   & = & \delta(p_1-b_1)\delta(q_2-a_2)\gamma(q_1,p_2)+
   \delta(p_1-b'_1)\delta(q_2-a'_2)\gamma'(q_1,p_2)\,, \label{II.7.T} \\
   & = & \delta(p_1-b_1)\delta(p_2-b'_2)\eta(q_1,q_2)+
   \delta(p_1-b'_1)\delta(p_2-b_2)\eta'(q_1,q_2)\,. \label{II.7.U}
\end{eqnarray}
According to eqs.\rf{II.7.R} to \rf{II.7.T}
\<\begin{array}{rcl}
  \rho & = & v\,\delta(q_1-a_1)\delta(q_2-a_2)\delta(p_1-b_1)\delta(p_2-b_2)\\
  & & +v'\delta(q_1-a'_1)\delta(q_2-a'_2)\delta(p_1-b'_1)\delta(p_2-b'_2)\,,
\end{array} \label{II.8} \>
with $v\geq0$, $v'\geq0$, ($v+v'=1$). Clearly, eqs.\rf{II.7.U} and \rf{II.8}
are incompatible, which establishes the non existence of $\rho$, and settles
the classical four marginal problem.

	This however does not settle the quantum problem. Actually, the above
example obviously cannot be strictly realized through a wave function in
accordance with eqs.\rf{II.5}. More than that, this example is so ``twisted''
that, even after smoothing out the $\delta$ measures in eqs.\rf{II.6.R} to
\rf{II.6.U}, approaching it close enough through a wave function appears as
very difficult (if not impossible). Instead, we develop a new
mathematical tool.

\section{Phase space Bell inequalities}
\setcounter{equation}{0}

Consider any choice of functions $r(q_1,q_2)$, $s(q_1,p_2)$,
$t(p_1,q_2)$, and $u(p_1,p_2)$, obeying
\< A \leq r(q_1,q_2)+s(q_1,p_2)+t(p_1,q_2)+u(p_1,p_2) \leq B \qquad
   (\forall q_1,q_2,p_1,p_2), \label{II.9}
\>
Multiply by $\rho(\vec q, \vec p)$, integrate over phase space and use
positivity and normalization of $\rho(\vec q, \vec p)$. We deduce that
the (classical as well as quantum) four marginal problem cannot have a
solution unless
\<\begin{array}{rcccl}
  A & \leq & \intd dq_1dq_2\,r(q_1,q_2)R(q_1,q_2)
      +\intd dq_1dp_2\,s(q_1,p_2)S(q_1,p_2) & & \\
  & & +\intd dp_1dq_2\,t(p_1,q_2)T(p_1,q_2)
      +\intd dp_1dp_2\,u(p_1,p_2)U(p_1,p_2) & \leq & B\,.
\end{array} \label{II.10} \>
Here $R$, $S$, $T$ and $U$ are defined by eqs.\rf{II.5} in the quantum
case. It turns out that a particularly interesting choice is
\<\begin{array}{rcl}
  r(q_1,q_2) & = & \sgn F_1(q_1)\,\sgn F_2(q_2)\,, \\
  s(q_1,p_2) & = & \sgn F_1(q_1)\,\sgn G_2(p_2)\,, \\
  t(p_1,q_2) & = & \sgn G_1(p_1)\,\sgn F_2(q_2)\,, \\
  u(p_1,p_2) & = & -\sgn G_1(p_1)\,\sgn G_2(p_2)\,,
\end{array} \label{II.11} \>
with $A=-2$, $B=+2$ and with $F_1$, $F_2$, $G_1$ and $G_2$ arbitrary
non vanishing functions\footnote{Note that, with this choice, the sum
$r+s+t+u$ assumes only its two extremal values $A=-2$ and $B=+2$,
which makes it in a sense optimal.\label{footnote 1}}. Then the
inequalities \rf{II.10} become a phase space analogue of the Bell
inequalities for spin variables.

	The necessary conditions \rf{II.10} provide us with an alternative
proof that the classical problem does not always admit a solution. Indeed, it
is readily seen that they are violated for the marginals \rf{II.6.R} to
\rf{II.6.U} and functions $F$'s and $G$'s such that
\[\begin{array}{c}
  F_1(a_1),F_2(a_2),G_1(b_1),G_2(b_2) > 0\,, \\
  F_1(a'_1),F_2(a'_2),G_1(b'_1),G_2(b'_2) < 0\,.
\end{array}\]

	We shall see in the next section that the necessary conditions
\rf{II.10} can be violated also in the quantum case. There, the analogy
between our correlation inequalities \rf{II.10} (with the choice \rf{II.11})
and Bell inequalities will become more apparent, especially as regards to
their implications.

\section{Solving the four marginal quantum problem}
\setcounter{equation}{0}

	This section is divided into four parts. In the first one, we prove
the existence of wave functions which violate the correlation inequalities
\rf{II.10}. Strictly speaking, this already settles the problem. However, the
explicit construction of such wave functions, which we present in subsections
B and C, is worthwhile in that it exhibits the physical implications of our
inequalities. In subsection D, we elaborate on the formal analogy with Bell
inequalities.

\subsection{Non constructive proof}

	One first notices that $\chi_1(q_1) \equiv {1\over2} \left [1+\sgn
F_1(q_1) \right]$ is the characteristic function of some set $S_1\subset \R$,
and similarly for $F_2$, $G_1$ and $G_2$, so that eqs.\rf{II.11} read
\<\begin{array}{rcl}
  r(q_1,q_2) & = & (2\chi_1-1)(2\chi_2-1)\,, \\
  s(q_1,p_2) & = & (2\chi_1-1)(2\chi'_2-1)\,, \\
  t(p_1,q_2) & = & (2\chi'_1-1)(2\chi_2-1)\,, \\
  u(p_1,p_2) & = & -(2\chi'_1-1)(2\chi'_2-1)\,,
\end{array} \label{III.1} \>
where $\chi_i$ stands for $\chi_i(q_i)$ and $\chi'_i$ for $\chi'_i(p_i)$,
($i=1,2$). Inequalities \rf{II.9} then become
\< 0 \leq \P \leq 1 \,,
\>
and in fact $\P(1-\P)=0$ (see footnote \ref{footnote 1}), where
$\P(q_1,q_2,p_1,p_2)$ is given by
\< \P = \chi_1+\chi_2+\chi'_1\chi'_2-\chi_1\chi_2-\chi_1\chi'_2-\chi'_1\chi_2 
   \,.
\>
Let us define a corresponding quantum operator $\Ph$ by
\< \Ph = \kh_1+\kh_2+\kh'_1\kh'_2-\kh_1\kh_2-\kh_1\kh'_2-\kh'_1\kh_2 \,,
   \label{III.4}
\>
where
\<\begin{array}{rcl}
  \kh_1 & = & \intd_{S_1}dq_1\,|q_1\rangle\langle q_1|\,\otimes
  \,{\bf 1}_2\,,\\
  \kh_2 & = & {\bf 1}_1\,\otimes\,\intd_{S_2}dq_2\,|q_2\rangle
  \langle q_2|\,,\\
  \kh'_1 & = & \intd_{S'_1}dp_1\,|p_1\rangle\langle p_1|\,\otimes
  \,{\bf 1}_2 \,,\\
  \kh'_2 & = & {\bf 1}_1\,\otimes\,\intd_{S'_2}dp_2\,|p_2\rangle
  \langle p_2|\,.
\end{array} \label{III.5} \>
The $\kh$'s are orthogonal projectors\footnote{In eqs.\rf{III.5}, $S_1$,
$S_2$, $S'_1$, $S'_2$ are the supports of $\kh_1$, $\kh_2$, $\kh'_1$,
$\kh'_2$ respectively. Also, $\int_{S_1}dq_1\,|q_1\rangle\langle q_1|$ is, in
standard Dirac notation, the orthogonal projection
$\psi(q_1)\rightarrow\chi_1(q_1)\psi(q_1)$, whereas
$\int_{S'_1}dp_1\,|p_1\rangle\langle p_1|$ is the orthogonal projection
$\tilde{\psi}(p_1)\rightarrow\chi'_1(p_1) \tilde{\psi}(p_1)$,
$\tilde{\psi}(p_1)$ being the Fourier transform of $\psi(q_1)$, and so on.}
($\kh\dagger=\kh$,\ $\kh^2=\kh$) acting on ${\cal H}\equiv
L^2(\R,dq_1)\otimes L^2(\R,dq_2)$. The product of two of them involving
different indices commutes, so that $\Ph$ is a (bounded) self-adjoint
operator.

	The inequalities \rf{II.10} to be tested in the quantum context then
become, for pure states $\rhoh=|\Psi\rangle\langle\Psi|$,
\< 0\leq\langle\Psi|\Ph|\Psi\rangle\leq1\qquad\forall\ |\Psi\rangle
   \in{\cal H}\mbox{ with }\langle\Psi|\Psi\rangle=1\,, \label{III.6}
\>
or, equivalently,
\< \Ph\geq0 \mbox{\ \ \ and\ \ \ } {\bf 1}-\Ph\geq0
   \mbox{\ \ \ in the operator sense}. \label{III.7}
\>

	Because $\kh_j$ fails to commute with $\kh'_j$ ($j=1,2$), $\Ph$ is
{\sl not} an orthogonal projector (see below), in contrast to the classical
equality $\P^2=\P$. Exploiting this fact leads to the
\newtheorem{proposition}{Proposition}
\begin{proposition}
  The operators $\Ph$ and (${\bf 1}-\Ph$) cannot be both positive.
  \label{proposition}
\end{proposition}
As a consequence, there is at least one $|\Psi\rangle\neq0$ such that the
inequalities $\langle\Psi|\Ph|\Psi\rangle\geq0$ and
$\langle\Psi|({\bf1}-\Ph)|\Psi\rangle\geq0$ cannot be simultaneously true.
This just means that one of the two inequalities \rf{III.6} is violated for
that $|\Psi\rangle$, which settles the question.

\noindent{\sl Proof of proposition 1:}

\noindent Assume that $\Ph$ and (${\bf 1}-\Ph$) are both positive. This would
imply
\< \Ph({\bf 1}-\Ph) \geq 0 \,, \label{III.9}
\>
(remember that the product of two positive {\sl commuting} operators is
positive). 

\noindent Now, a straightforward calculation of $\Ph^2$ from eq.\rf{III.4}
yields
\< \Ph^2 = \Ph-\left[\kh_1,\kh'_1\right] \left[\kh_2,\kh'_2\right]\,,
\>
and eq.\rf{III.9} would mean that $\left[\kh_1,\kh'_1\right]
\left[\kh_2,\kh'_2\right]$ is a positive operator. That this is wrong is not
surprising. Let us show it. Take a factorized $|\Psi\rangle$, namely
$|\Psi\rangle=|\Phi_1\rangle\otimes|\Phi_2\rangle$, so that
\[ \langle\Psi|\Ph({\bf 1}-\Ph)|\Psi\rangle = 
   -\langle\Phi_1|i\left[\kh_1,\kh'_1\right]|\Phi_1\rangle
   \langle\Phi_2|i\left[\kh_2,\kh'_2\right]|\Phi_2\rangle\,.
\]
It is enough to show that, for a given choice of the characteristic functions
$\chi$ and $\chi'$, the real number
\< R[\Phi] \equiv \langle\Phi|i\left[\kh,\kh'\right]|\Phi\rangle
\>
can assume both signs when $|\Phi\rangle$ is varied.

	Let us define
\[ |\Phi^+\rangle = \kh\,|\Phi\rangle\,, \qquad
   |\Phi^-\rangle = ({\bf 1}-\kh)|\Phi\rangle\,.
\]
Using the identity
\[ \left[\kh,\kh'\right] = \kh\kh'({\bf 1}-\kh)-({\bf 1}-\kh)\kh'\kh
\]
gives $R[\Phi]$ the form
\[ R[\Phi] = i\langle\Phi^+|\kh'|\Phi^-\rangle
   -i\langle\Phi^-|\kh'|\Phi^+\rangle\,.
\]
Obviously, for
$|\widetilde{\Phi}\rangle=|\Phi^+\rangle-|\Phi^-\rangle$, one has
$R[\widetilde{\Phi}]=-R[\Phi]$.

	This concludes the proof.

\noindent {\sl Remarks:}

\noindent 1) When the wave function $|\Psi\rangle$ factorizes, i.e.\ 
$\Psi(q_1,q_2) = \Phi_1(q_1)\Phi_2(q_2)$, a corresponding probability
distribution $\rho$ always exists, namely
\< \rho(q_1,q_2,p_1,p_2) = |\Phi_1(q_1)|^2\,|\Phi_2(q_2)|^2\,
   |\tilde{\Phi}_1(p_1)|^2\,|\tilde{\Phi}_2(p_2)|^2 \,, \label{factor}
\>
where the $\tilde{\Phi}_i$'s are the Fourier transforms
\< \tilde{\Phi}_i(p_i) = {1\over\sqrt{2\pi}} \int_{-\infty}^{+\infty}dq_i\,
   \mbox{e}^{-ip_iq_i}\,\Phi_i(q_i)\,, \qquad (i=1,2).
\>
Of course, this implies that eqs.\rf{III.6} are automatically satisfied for
such factorized $|\Psi\rangle$'s (which can also be checked from
eq.\rf{III.4}).

\noindent 2) The fact (used in the proof) that
$\langle\Psi|\Ph({\bf 1}-\Ph)|\Psi\rangle<0$ for some factorized
$|\Psi\rangle$'s is {\sl not} inconsistent with the inequalities
$0\leq\langle\Psi|\Ph|\Psi\rangle\leq1$ which are satisfied for those
$|\Psi\rangle$'s.

\subsection{Construction}

	We want to find wave functions $|\Psi\rangle$ violating the
inequalities \rf{III.6}. According to the first of the above remarks, one has
to depart from the class of factorized $|\Psi\rangle$'s. The simplest way to
do it is to take just a sum of two such products.

	Choose first
\[ S_1=S_2\equiv S \qquad \mbox{ and } \qquad S'_1=S'_2\equiv S'\,,
\]
so that
\< \Ph = \kh\otimes{\bf 1}_2+{\bf 1}_1\otimes\kh+\kh'\otimes\kh'-\kh\otimes\kh
   -\kh\otimes\kh'-\kh'\otimes\kh \,. \label{III.12}
\>
Take next
\<\left\{\begin{array}{l}
  |\Psi\rangle={1\over \sqrt{1+|\lambda|^2}}
    (|\phi\rangle+\lambda|\varphi\rangle)
    \qquad (\lambda\in\C) \,, \\
  \mbox{with } \quad \left\{ \begin{array}{l}
    |\phi\rangle=|\phi_1\rangle\otimes|\phi_2\rangle\,, \qquad
    |\varphi\rangle=|\varphi_1\rangle\otimes|\varphi_2\rangle\,, \\
    \langle\phi_1|\phi_1\rangle=\langle\phi_2|\phi_2\rangle=
    \langle\varphi_1|\varphi_1\rangle=\langle\varphi_2|\varphi_2\rangle=1,
    \quad \langle\phi_1|\varphi_1\rangle=0\,,
    \end{array} \right.
\end{array} \right. \label{III.13} \>
so that $|\Psi\rangle$ is properly normalized.

	For the moment, choose also
\< \phi_1=\phi_2\equiv f\quad \mbox{ and } \quad\varphi_1=\varphi_2\equiv g\,.
   \label{III.14}
\>
with
\< \langle f|f\rangle=\langle g|g\rangle=1\,, \quad \langle f|g\rangle=0\,.
   \label{III.15}
\>
Then
\< \langle\Psi|\Ph|\Psi\rangle = {1\over{1+|\lambda|^2}} \left[
     \langle\phi|\Ph|\phi\rangle +(\lambda\langle\phi|\Ph|\varphi\rangle+c.c.)
     +|\lambda|^2\langle\varphi|\Ph|\varphi\rangle \right]\,, \label{III.16}
\>
with
\<\begin{array}{rcl}
  \langle\phi|\Ph|\phi\rangle & = & 2\langle f|\kh|f\rangle
    +\langle f|\kh'|f\rangle^2 -\langle f|\kh|f\rangle^2
    -2\langle f|\kh|f\rangle\langle f|\kh'|f\rangle \,, \\
  \langle\varphi|\Ph|\varphi\rangle & = & 2\langle g|\kh|g\rangle
    +\langle g|\kh'|g\rangle^2 -\langle g|\kh|g\rangle^2
    -2\langle g|\kh|g\rangle\langle g|\kh'|g\rangle \,, \\
  \langle\phi|\Ph|\varphi\rangle & = & \langle f|\kh'|g\rangle^2
    -\langle f|\kh|g\rangle^2
    -2\langle f|\kh|g\rangle\langle f|\kh'|g\rangle\,.
\end{array} \label{III.17} \>
We already know that $0\leq\langle\phi|\Ph|\phi\rangle\leq1$ and
$0\leq\langle\varphi|\Ph|\varphi\rangle\leq1$\,. Clearly, in view of
\rf{III.16}, our goal will be reached (namely $\langle\Psi|\Ph|\Psi\rangle
<0$ or $\langle\Psi|\Ph|\Psi\rangle>1$) if one can find $f$ and $g$ such that
\< |\langle\phi|\Ph|\varphi\rangle|^2 >
   \langle\phi|\Ph|\phi\rangle\langle\varphi|\Ph|\varphi\rangle\,.
   \label{III.18}
\>
We claim that this can be achieved with $S=S'=(0,\infty)$, $f(q)\equiv
\langle q|f\rangle$ an even, normalized function in $L^2(-\infty,\infty)$, and
\< g(q)\equiv\langle q|g\rangle = \sgn(q)\,f(q)\,. \label{III.19}
\>
With this choice, eqs.\rf{III.15} are automatically satisfied and
\< \langle f|\kh|f\rangle = \langle g|\kh|g\rangle =
   \langle f|\kh|g\rangle = {1\over 2}\,. \label{III.20}
\>
Also, since the Fourier transforms $\tilde{f}(p)$ and $\tilde{g}(p)$ are
respectively even and odd functions
\< \langle f|\kh'|f\rangle = \langle g|\kh'|g\rangle = {1\over 2}\,.
   \label{III.21}
\>
As for the non trivial interference term $\langle f|\kh'|g\rangle$, it is
given by (see appendix A)
\< \langle f|\kh'|g\rangle = -{i\over\pi} \int_0^\infty dq\int_0^\infty dq'
   \,f^*(q)f(q')\left({1\over q+q'}-{P\over q-q'}\right)\,. \label{III.22}
\>
At this stage, it is advantageous to take $f$ as a {\sl real} function, so
that by symmetry
\[ \langle f|\kh'|g\rangle = -{i\over\pi} \int_0^\infty dq\int_0^\infty dq'
   \,{f(q)f(q')\over q+q'}\,.
\]
Let us set
\begin{eqnarray}
  h(q) & = & \sqrt{2}\,f(q)\,, \label{III.23} \\
  K(q,q') & = & {1\over\pi}\,{1\over q+q'}\,. \label{III.24}
\end{eqnarray}
Then
\< \langle f|\kh'|g\rangle = -{i\over 2}\,\gamma\,,
   \qquad\qquad (\gamma\in\R) \label{III.25}
\>
with
\< \gamma = \int_0^\infty dq\int_0^\infty dq'\,h(q)\,K(q,q')\,h(q')\,,
   \label{III.26}
\>
and
\< \|h\|_{L^2(0,\infty)} = 1\,. \label{III.27}
\>
The insertion of eqs.\rf{III.20}, \rf{III.21} and \rf{III.25} in \rf{III.17}
gives
\[\begin{array}{c}
  \langle\phi|\Ph|\phi\rangle = \langle\varphi|\Ph|\varphi\rangle
    = {1\over 2}\,, \\
  \langle\phi|\Ph|\varphi\rangle = -{1\over 4}(1+\gamma^2)
    +{\displaystyle i\over 2}\,\gamma\,, 
\end{array} \]
so that eq.\rf{III.18} reads
\[ (\gamma^2+1)^2+4\gamma^2 > 4 \,,
\]
which is satisfied provided that
\< |\gamma| > \sqrt{2\sqrt{3}-3} \,\cong\, 0.6813 \label{III.28}
\>
Moreover, with $\lambda=\rho\,\e{i\theta}$, eq.\rf{III.16} becomes
\< \langle\Psi|\Ph|\Psi\rangle = {1\over 2}-{\rho\over 2(1+\rho^2)}
   \left[(1+\gamma^2)\cos\theta+2\gamma\sin\theta\right]\,. \label{III.29}
\>
We already know that $|\gamma|$ cannot exceed 1, because
$|\langle f|\kh'|g\rangle|^2 \leq \langle f|\kh'|f\rangle
\langle g|\kh'|g\rangle={1\over4}$. Are there however some $h$'s (subjected
to \rf{III.27}) such that $\gamma$ (given by eq.\rf{III.26}) fulfils
\rf{III.28}? If this occurs we have reached our goal and it only remains to
maximize $|\gamma|$ in order to obtain the extremal values of
$\langle\Psi|\Ph|\Psi\rangle$ (within the present scheme) through
eq.\rf{III.29}. In other words, one has to solve the problem
\renewcommand{\arraystretch}{.7}
\[ \gamma_0 \equiv \build{\sup}
   {\begin{array}{c}
   \scriptstyle \|h\|_{\scriptscriptstyle L^2(0,\infty)}=1 \\
   \scriptstyle h=h^* \end{array}}
   |\langle h|K|h\rangle| = \ ?
\]
\renewcommand{\arraystretch}{1.2}
\noindent In appendix B, it is shown that the (bounded) integral operator $K$
with kernel \rf{III.24} on $L^2(0,\infty)$ is positive and has the purely
continuous spectrum $[0,1]$. This immediately entails $\gamma_0=1$, and we
get
\[ \left.\langle\Psi|\Ph|\Psi\rangle\right|_{\gamma=1} = 
   {1\over 2}-{\rho\over 1+\rho^2}(\cos\theta+\sin\theta)\,,
\]
\<\begin{array}{rclclcl}
  \build{\inf}{\lambda} \left.\langle\Psi|\Ph|\Psi\rangle\right|_{\gamma=1}
  & = & \left.\langle\Psi|\Ph|\Psi\rangle\right|_{\gamma=\rho=1,
  \theta={\pi\over4}} & = & {1-\sqrt{2}\over 2 } & \cong & -0.2071 \\
  \build{\sup}{\lambda} \left.\langle\Psi|\Ph|\Psi\rangle\right|_{\gamma=1}
  & = & \left.\langle\Psi|\Ph|\Psi\rangle\right|_{\gamma=\rho=1,
  \theta={-3\pi\over4}} & = & {1+\sqrt{2}\over 2 } & \cong & 1.2071
\end{array} \label{III.30} \>
Actually, as discussed in appendix B, due to the continuous spectrum of $K$,
these extremal values cannot be strictly reached, but only approached
arbitrarily close via a family of normalized functions $h$, e.g.
\[ h_L(q) = {\theta(L-q)\over \sqrt{\ln(L+1)}}{1\over \sqrt{q+1}},
   \qquad L\rightarrow\infty
\]
or smoothed forms of this. Of course, other functions $h$ will also do the
job (although less perfectly), that is meet the crucial requirement
\rf{III.28}. Taking for example $h(q)={1\over q+1}$ (which is normalized in
$L^2(0,\infty)$), one gets
\[ \gamma = {\pi\over 4} \cong 0.7854
\]
Finally, collecting the equations \rf{III.13}, \rf{III.14}, \rf{III.19} and
\rf{III.23}, together with $\lambda={\pi\over4},-{3\pi\over4}$, one obtains
the wave functions leading to the maximal violations \rf{III.30}
\< \Psi_{\pm}(q_1,q_2) = {1\over 2\sqrt{2}}\left[1\pm
   \e{i{\pi\over 4}}\sgn(q_1)\sgn(q_2)\right]h(|q_1|)h(|q_2|)\,,
   \label{III.31}
\>
where $h(q)$ stands for some regularized form of $1\over\sqrt{q}$, with
$\int_0^\infty dq\,h(q)^2=1$.

\subsection{Introducing Einstein locality and relative motion}

	Let us now interpret $q_1$ and $q_2$ as the coordinates of two
particles (rather than the $x$ and $y$ coordinates of the same particle)..
Then the wave functions \rf{III.31} describe states of two particles not
spatially separated and with zero relative momentum. These two restrictions
can be easily disposed of.

	First, it can be checked that nothing is essentially changed in the
previous derivation if one keeps
\< S_1=S'_1=(0,\infty)\,; \qquad \phi_1(q_1)=f(q_1)\,, \quad
    \varphi_1(q_1) = \sgn(q_1) f(q_1)\,, \label{III.32}
\>
but replaces
\< S_2=S'_2=(0,\infty)\,; \qquad \phi_2(q_2)=f(q_2)\,, \quad
    \varphi_2(q_2) = \sgn(q_2) f(q_2)\,, \label{III.33}
\>
by
\[ S_2=(a,\infty)\,, \quad S'_2=(0,\infty)\,; \qquad
   \phi_2(q_2)=f(q_2-a)\,, \quad \varphi_2(q_2) = \sgn(q_2-a) f(q_2-a)\,.
\]
Then eq.\rf{III.31} becomes
\[ \Psi_{\pm}(q_1,q_2) = {1\over 2\sqrt{2}}\left[1\pm
  \e{i{\pi\over 4}}\sgn(q_1)\sgn(q_2-a)\right]h(|q_1|)h(|q_2-a|)\,,
\]
with $a$ arbitrary.

	This allows us to let Einstein locality enter the game.

	Similarly, nothing is essentially changed if one keeps eqs.\rf{III.32}
but replaces eqs.\rf{III.33} by
\[ S_2=(0,\infty)\,, \quad S'_2=(P,\infty)\,; \qquad
   \phi_2(q_2)=\e{iPq_2}f(q_2)\,,
   \quad \varphi_2(q_2) = \e{iPq_2}\sgn(q_2) f(q_2)\,.
\]
Then eq.\rf{III.31} becomes
\[ \Psi_{\pm}(q_1,q_2) = {1\over 2\sqrt{2}}\left[1\pm
   \e{i{\pi\over 4}}\sgn(q_1)\sgn(q_2)\right]
   \e{iPq_2}h(|q_1|)h(|q_2|)\,,
\]
with $P$ arbitrary.

	This allows us to put the two particles in relative motion.

\subsection{Analogy with Bell spin $1\over2$ correlation inequalities}

	Let us denote by $|+\rangle$ a normalized function $f$ close to the
(symmetrized) eigenfunction of the operator $K$ with ``eigenvalue''
$\lambda_0=1$ (i.e.\ $\gamma\cong1$ in eq.\rf{III.25}), and by $|-\rangle$ the
orthogonal function $g$ (as given by eq.\rf{III.19}). Consider the subspace
$V=\mbox{span}(|+\rangle,|-\rangle)$ of the full 1-particle Hilbert space,
together with the orthogonal projector $\Pi$ onto $V$. Call $\Gamma$
(resp. $\Gamma'$) the restriction of $\kh$ (resp. $\kh'$) to the 2-dimensional
space $V$
\[ \Gamma=\Pi\,\kh\,\Pi\,, \qquad \Gamma'=\Pi\,\kh'\,\Pi\,.
\]
Then eqs.\rf{III.20}, \rf{III.21} and \rf{III.25} tell us that $\Gamma$ and
$\Gamma'$ are represented in the orthonormal basis $\{|+\rangle,|-\rangle\}$
by the matrices
\[\begin{array}{rcl}
  \Gamma = \left(\begin{array}{lr} {1\over2} & {1\over2} \\
  {1\over2} & {1\over2} \end{array}\right) & = & {1\over2}\,(1+\sigma_x)\,,\\ 
  [.2in] \Gamma' = \left(\begin{array}{cc} {1\over2} & {i\over2}\gamma \\
  -{i\over2}\gamma & {1\over2} \end{array}\right) & = &
  {1\over2}\,(1-\gamma\sigma_y)\,.
\end{array}\]
In the idealized limit $\gamma\rightarrow1$ (and only in this limit), one
observes that $\Gamma$ and $\Gamma'$ are themselves orthogonal projections
$V\rightarrow V$
\[\begin{array}{rclrcl} \Gamma&=&\Gamma^\dagger\,, & \Gamma^2&=&\Gamma\,, \\
  \Gamma'&=&\Gamma'^\dagger\,,\qquad & \Gamma'^2&=&\Gamma'\,.
\end{array}\]
This implies that both operators $\kh$ and $\kh'$ leave the subspace $V$
invariant
\[ [\Pi,\kh] = [\Pi,\kh'] = 0\,.
\]
Indeed, a straightforward calculation shows that
\[ [(\mbox{\bf 1}-\Pi)\kh\Pi]^\dagger[(\mbox{\bf 1}-\Pi)\kh\Pi] = 0\,,
\]
which entails $(\mbox{\bf 1}-\Pi)\kh\Pi=0$ and $\kh\Pi=\Pi\kh$. The same for
$\kh'$.

Hence, in the 2-particle Hilbert space, the operator \rf{III.12} also leaves invariant $V\otimes V$, and $\Pb:=\Ph\Pi$ assumes the simple form
\< \Pb = {1\over2}+{1\over4}\left(\sigma_y^{(1)}\sigma_y^{(2)}
   -\sigma_x^{(1)}\sigma_x^{(2)}+\sigma_x^{(1)}\sigma_y^{(2)}
   +\sigma_y^{(1)}\sigma_x^{(2)}\right)\,, \label{III.34}
\>
whereas the maximally violating wave functions \rf{III.31} read
\< |\Psi_\pm\rangle = {1\over\sqrt{2}}\left(|+\rangle^{(1)}|+\rangle^{(2)}
   \pm\e{i{\pi\over4}}|-\rangle^{(1)}|-\rangle^{(2)}\right)\,.
   \label{III.35}
\>
From eq.\rf{III.34} one can check that
\< \Pb({\bf 1}-\Pb) = -{1\over4}\,\sigma_z^{(1)}\sigma_z^{(2)}\,,
   \label{III.36}
\>
which is just the projected form of $\Ph({\bf 1}-\Ph)=[\kh_1,\kh'_1]\,
[\kh_2,\kh'_2]$, and the expectation value of the operator \rf{III.36} is
\[ \langle\Psi_\pm|\Pb({\bf 1}-\Pb)|\Psi_\pm\rangle = -{1\over4}
\]
for the wave functions \rf{III.35}.

	The result \rf{III.30} is also directly recovered from
eqs.\rf{III.34} and \rf{III.35}
\[ \langle\Psi_\pm|\Pb|\Psi_\pm\rangle = {1\mp\sqrt{2}\over2}\,.
\]

	Then one sees that, in the idealized limit $\gamma\rightarrow1$, the
original phase space setting up of the problem is formally equivalent to the
standard EPR setting up for a two spin $1\over2$ system, together with its
classical Bell inequalities.

\section{General solution of the three marginal problem}
\setcounter{equation}{0}

We have proved here the impossibility of reproducing quantum
probabilities of four CCS as marginals. Roy and Singh \cite{RS7} have
given examples to show that reproducing three CCS is possible. In this
section, we construct the most general nonnegative phase space density
which reproduces three different (noncommuting) CCS as marginals. Our
results encapsulate the extent to which noncommuting CCS can be
simultaneously realized in quantum mechanics.

\indent Among the four marginals $R,S,T,U$ obeying the compatibility
conditions \rf{II.4} which are at our disposal, the particular choice of
three of them is completely irrelevant. For definiteness, we choose $R,T$ and
$U$, which we rename $\sigma_0(q_1,q_2)$, $\sigma_1(p_1,q_2)$ and
$\sigma_2(p_1,p_2)$.

	We assume that these marginals are probability densities in the full
mathematical sense, that is they are true (integrable and non negative) {\sl
functions}. This means that we restrict our marginal probability
distributions to {\sl absolutely continuous} measures (with respect to
Lebesgue measure) in $\R^2$. Notice that such a restriction is automatic in
the quantum case, due to eqs.\rf{II.5}.

	Likewise, we look for the general solution of the three marginal
problem in the class of absolutely continuous measures in the phase space
$\R^4$. This means that we want to describe all the solutions $\rho$ of the
equations
\<\begin{array}{c}
  \displaystyle \sigma_0(q_1,q_2)=\int dp_1dp_2\,\rho(\qp)\,,\\
  \displaystyle \sigma_1(p_1,q_2)=\int dq_1dp_2\,\rho(\qp)\,,\\
  \displaystyle \sigma_2(p_1,p_2)=\int dq_1dq_2\,\rho(\qp)\,,
\end{array} \label{IV.1} \>
which belong to $L^1(\R^4,d^2q\,d^2p)$.

	Notice that this is a restricted problem even in the quantum case,
since nothing prevents a probability measure containing a singular part to
project on marginals which are $L^1$-functions. To some extent, the above
restrictions can be removed, allowing us to include e.g. probability measures
partly concentrated on submanifolds of the phase space. However, dealing with
such extensions at some degree of generality requires painful manipulations,
and we shall ignore them here\footnote{Special cases are treated in
\cite{RS5} and \cite{RS7}.}. As for the full inclusion of singular
measures, it appears as both delicate and of little practical interest.

Let us introduce the one variable marginals
\<\begin{array}{c}
   \displaystyle \sigma_{01}(q_2) = \int dq_1\,\sigma_0(q_1,q_2)\,, \\
   \displaystyle \sigma_{12}(p_1) = \int dq_2\,\sigma_1(p_1,q_2)\,.
\end{array}\label{IV.4}
\>
Owing to the compatibility conditions \rf{II.4}, these definitions are
equivalent to
\<\begin{array}{c}
   \displaystyle \sigma_{01}(q_2) = \int dp_1\,\sigma_1(p_1,q_2)\,, \\
   \displaystyle \sigma_{12}(p_1) = \int dp_2\,\sigma_2(p_1,p_2)\,.
\end{array}\label{IV.5}
\>

	As the support properties of the functions $\sigma_j$ (which are
allowed to vanish on some parts of $\R^2$) are not innocent in the
forthcoming construction, we need to pay attention to them. Let
$\Sigma_j\subset\R^2$ ($j=0,1,2$) be the essential support of $\sigma_j$. The
above compatibility conditions, together with the positivity conditions
$\sigma_j\geq0$, clearly yield two constraints on the supports $\Sigma_j$,
namely
\<\begin{array}{r}
  \{q_2\in\R\ |\ \exists\ q_1\in\R \mbox{\ such that }
    (q_1,q_2)\in\Sigma_0\} \quad = \qquad\qquad\qquad\qquad \\
  \{q_2\in\R\,|\,\exists\ p_1\in\R \mbox{\ such that }
    (p_1,q_2)\in\Sigma_1\}\,, \\
  \{p_1\in\R\,|\,\exists\ q_2\in\R \mbox{\ such that }
    (p_1,q_2)\in\Sigma_1\} \quad = \qquad\qquad\qquad\qquad \\
  \{p_1\in\R\,|\,\exists\ p_2\in\R \mbox{\ such that }
    (p_1,p_2)\in\Sigma_2\}\,.
\end{array}\label{IV.6}
\>
To the $\Sigma_j$'s we associate the subsets $E_j$'s of the phase space
defined by
\<\begin{array}{c}
  E_0 = \{\qp\ |\ (q_1,q_2)\in\Sigma_0\,, (p_1,p_2)\in\R^2 \}\,, \\
  E_1 = \{\qp\ |\ (p_1,q_2)\in\Sigma_1\,, (q_1,p_2)\in\R^2 \}\,, \\
  E_2 = \{\qp\ |\ (p_1,p_2)\in\Sigma_2\,, (q_1,q_2)\in\R^2 \}\,.
\end{array}\label{IV.7}
\>
Finally, we denote by $E$ the intersection of the $E_j$'s
\< E = E_0\cap E_1\cap E_2\,. \label{IV.8}
\>
Clearly, due to positivity again, any solution $\rho$ of eqs.\rf{IV.1} must
have its essential support contained in $E$.

	The three marginal problem in the precise form stated above is then completely solved by
\newtheorem{theorem}{Theorem}
\begin{theorem}

1) The Lebesgue measure of $E$ is not zero and the function $\rho_0$ defined
(a.e.) by
\< \rho_0(\qp) = \left\{ \begin{array}{l}
     \sigma_0(q_1,q_2) {\displaystyle 1\over\displaystyle\sigma_{01}(q_2)}
     \sigma_1(p_1,q_2) {\displaystyle1\over\displaystyle\sigma_{12}(p_1)}
     \sigma_2(p_1,p_2) \mbox{\ \ if\ }(\qp)\in E\,, \\
     0 \qquad \mbox{ otherwise,} \end{array}\right. \label{IV.9}
\>
is a non negative solution of the problem \rf{IV.1} in $L^1(\R^4,d^2q\,d^2p)$.

2) The general solution $\rho$ of \rf{IV.1} in $L^1(\R^4,d^2q\,d^2p)$ is given
by
\< \rho(\qp) = \rho_0(\qp)+\lambda\,\Delta(\qp)\,, \label{IV.10}
\>
where
\< \lambda\in\left[-{1\over m_+},{1\over m_-}\right]\,, \label{IV.11}
\>
and
\renewcommand{\arraystretch}{1.9}
\<\begin{array}{l}
  \Delta(\qp) = F(\qp)-\rho_0(\qp)
  \left[\displaystyle
  {1\over\sigma_0(q_1,q_2)}\intd dp'_1dp'_2
  \,F(q_1,q_2,p'_1,p'_2) \right. \\
   \qquad +\displaystyle
  {1\over\sigma_1(p_1,q_2)}\intd dq'_1dp'_2
  \,F(q'_1,q_2,p_1,p'_2) +
  {1\over\sigma_2(p_1,p_2)}\intd dq'_1dq'_2
  \,F(q'_1,q'_2,p_1,p_2) \\
   \qquad\left. -\displaystyle
  {1\over\sigma_{01}(q_2)}\intd dq'_1dp'_1dp'_2
  \,F(q'_1,q_2,p'_1,p'_2) -
  {1\over\sigma_{12}(p_1)}\intd dq'_1dq'_2dp'_2
  \,F(q'_1,q'_2,p_1,p'_2) \right],
  \end{array}\label{IV.12}
\>
\renewcommand{\arraystretch}{1.4}
$F$ being an {\sl arbitrary} $L^1(\R^4,d^2q\,d^2p)$-function with essential
support contained in $E$. The (F-dependent) constants $m_\pm$ in \rf{IV.11}
are defined as
\< m_+ = \ \build{\mbox{ ess sup }}{(\overrightarrow{\scriptstyle q},
   \overrightarrow{\scriptstyle p})\in E} \ {\Delta(\qp)\over\rho_0(\qp)}\,,
   \qquad
   m_- = \ -\build{\mbox{ ess inf }}{(\overrightarrow{\scriptstyle q},
   \overrightarrow{\scriptstyle p})\in E} \ {\Delta(\qp)\over\rho_0(\qp)}\,,
   \label{IV.13}
\>
and are both positive if $\Delta\not=0$ ($m_+=\infty$ or/and $m_-=\infty$ are
not excluded). 
\label{theorem}
\end{theorem}
\noindent{\sl Proof:}

\noindent 1) To begin with, $\rho_0$ given by \rf{IV.9} is well defined and
non negative. Indeed, due to \rf{IV.4} (or \rf{IV.5}) and the positivity of
the $\sigma_j$'s, $\sigma_{01}(q_2)$ and $\sigma_{12}(p_1)$ are a.e. non zero
for $(\qp)\in E_0$ and $E_1$ (or $E_1$ and $E_2$), so that the denominators in
eq.\rf{IV.9} do not vanish on $E$ (except maybe on sets of Lebesgue measure
0).

	Next, in order to check that $\rho_0$ obeys the first equation
\rf{IV.1}, we consider the integral
\< \int dp_1 \int dp_2\,\rho_0(\qp) \label{IV.14}
\>
with this specific order of the $p$ integrations. According to the relations
\rf{IV.6} and the definition of $E$, one observes first that the projection
of $E$ on the $(p_1,p_2)$ plane is the set $\Sigma_2$, so that the
integration over $p_2$ removes the factor $\sigma_2/\sigma_{12}$ in $\rho_0$;
and second, that the projections of $\Sigma_1$ and $\Sigma_2$ on $p_1$
coincide, so that the integration over $p_1$ removes the factor
$\sigma_1/\sigma_{01}$ in $\rho_0$, and one is left with the expected result
$\sigma_0(q_1,q_2)$. We can now write
\< \int dp_1dp_2\,\rho_0(\qp) = \sigma_0(q_1,q_2) \label{IV.15}
\>
where, thanks to Fubini theorem, the integration order is completely
irrelevant. The other two equations \rf{IV.1} are derived in a similar way.

	This calculation shows at once that the Lebesgue measure of $E$ is
not zero and that $\rho_0\in L^1(\R^4,d^2q\,d^2p)$.

\noindent 2) That any non negative solution $\rho$ of eqs.\rf{IV.1} admits
the representation \rf{IV.10}-\rf{IV.12} is easy to establish. Indeed, since
the essential support of $\rho$ is necessarily contained in $E$, we are
allowed to take $F=\rho$ in eq.\rf{IV.12}, which gives (using \rf{IV.1})
\< \Delta(\qp) = \rho(\qp)-\rho_0(\qp)\,. \label{IV.16}
\>
Then, from \rf{IV.13}
\[ m_- = -\build{\mbox{ ess inf }}{(\overrightarrow{\scriptstyle q},
   \overrightarrow{\scriptstyle p})\in E}
   \left({\rho\over\rho_0}-1\right) \leq 1\,.
\]
As $1/m_-\geq 1$ in \rf{IV.11}, we can choose $\lambda=1$, which makes
eq.\rf{IV.16} equivalent to the representation \rf{IV.10}.

	It remains to show that any function $\rho$ defined by
\rf{IV.10} to \rf{IV.13} (and thus with essential support $E$) is a non
negative solution of eqs.\rf{IV.1} in $L^1(\R^4,d^2q\,d^2p)$. In order to
prove that $\rho$ satisfies the first equation \rf{IV.1}, we rearrange
pairwise the right-hand side of \rf{IV.12} as follows
\begin{eqnarray}
  \Delta &=& \left[F-{\rho_0\over\sigma_0} \int dp'_1dp'_2\,F\right]
  -\left[{\rho_0\over\sigma_1} \int dq'_1dp'_2\,F
  - {\rho_0\over\sigma_{01}} \int dq'_1dp'_1dp'_2\,F\right] \nonumber \\
  && -\left[{\rho_0\over\sigma_2} \int dq'_1dq'_2\,F
  - {\rho_0\over\sigma_{12}} \int dq'_1dq'_2dp'_2\,F\right]\,. \label{IV.18}
\end{eqnarray}
Then, integrating the right-hand side over $p_1$ and $p_2$, one finds, by an
extensive use of eqs.\rf{IV.4} to \rf{IV.6} as in part 1), that the two terms
coming from each square bracket cancel each other, leading to
\[ \int dp_1dp_2\,\Delta(\qp) = 0\,.
\]
This, with \rf{IV.10} and \rf{IV.15}, implies that $\rho$ satisfies the first
equation \rf{IV.1}. That it satisfies the other two equations \rf{IV.1} is
proved in a similar way.

	This calculation also shows that $\rho\in L^1(\R^4,d^2q\,d^2p)$.

	Finally
\[ \int_E d^2qd^2p\,\Delta(\qp) = 0\,,
\]
which implies that $m_\pm$ in eqs.\rf{IV.13} are both strictly positive if
$\Delta$ does not vanish a.e. on $\R^4$. The positivity of $\rho$ is then a
trivial consequence of eqs.\rf{IV.10}, \rf{IV.11} and \rf{IV.13}.

	The proof is complete.

\noindent {\sl Remark:}

\noindent Theorem 1, as it is stated above, deals with $L^1$ functions, and
thus excludes the occurrence of Dirac measures. We insist on the fact that
this is unnecessarily restrictive. Indeed Dirac measures can be easily
accommodated and the theorem suitably rephrased, to the price however of
cumbersome mathematical intricacies which we do not want to enter into.

An immediate corollary of Proposition \ref{proposition} and Theorem
\ref{theorem} is
\begin{theorem}[Three marginal theorem]
Let R, S, T and U be probablilty distributions for $(q_1,q_2)$,
$(q_1,p_2)$, $(p_1,q_2)$ and $(p_1,p_2)$ obeying the consistency
conditions \rf{II.4}. Given $n$ arbitrary distributions among
$\{R,S,T,U\}$, a necessary and sufficient condition for them to be
marginals of a probability density in the $4$-dimensional phase space is
$n\leq3$.
\end{theorem}

\section{Conclusions}
\setcounter{equation}{0}

     We have solved the four marginal problem in four dimensional
phase space thus proving a long standing conjecture \cite{RS7} and
vastly improving the first results of Martin and Roy \cite{MartinRoy1}
which dealt with infinite number of marginals. To achieve this, we
first derived ``phase space Bell inequalities'' which have their own
interest. Actually they allow, at least in principle, direct
``experimental'' tests of the orthodox-versus-hidden variable
interpretations of quantum mechanics within the position-momentum
sector, analoguous to those performed within the spin sector.

The technique of phase space Bell inequalities established here has
applications to quantum information processing. Generalizing the 
example \rf{factor}, one can show that for any separable density
operator $\rho$ one can construct a phase space density obeying the
four marginal conditions. Hence, the Bell inequalities \rf{II.10},
with $R$, $S$, $T$ and $U$ given by \rf{II.5} must hold for every
separable quantum state, irrespective of any physical interpretation
of the associated phase space density. Their violation by a quantum
state is a signature and even a quantitative measure of entanglement
of this state.

 We have also
constructed the most general positive definite phase space density
which has the maximum number of marginals (three) coinciding with
corresponding quantum probabilities of three different (noncommuting)
CCS.  These results should be useful in the construction of maximally
realistic quantum theories.

\section{Acknowledgements}

     We thank Andr\'e Martin for collaboration in the initial stages of
this work.  One of us (SMR) thanks A. Fine and A. Garg for some
remarks on the three marginal problem many years ago.

\newpage\appendix

\def\thesection{Appendix \Alph{section}.}
\def\theequation{\Alph{section}.\arabic{equation}}

\section{Proof of equation \rf{III.22}}
\setcounter{equation}{0}

Since $S'=(0,\infty)$, one has
\[ \kh'(p)\tilde{g}(p) = \theta(p)\tilde{g}(p)\,.
\]
Assuming first that $g$ belongs to $\cal S$ (the Schwartz space of infinitely
differentiable functions on $\R$ with fast decrease at infinity), one can
write
\[ (\kh'g)(q) = \int_{-\infty}^\infty dq'\,\tilde{\theta}(q-q')g(q')\,,
\]
where $\tilde{\theta}$ is the Fourier transform of $\theta$ in the
distribution-theoretic sense
\[ \tilde{\theta}(q) \equiv {1\over 2\pi}\int_{-\infty}^\infty dp\,
   \e{ipq}\,\theta(p)
   = {i\over 2\pi}{P\over q}+{1\over 2}\,\delta(q)\,.
\]
Then, if $f$ also belongs to $\cal S$
\[ \langle f|\kh'|g\rangle = {i\over 2\pi} \int_{-\infty}^{\infty}dq\,f^*(q)
   \int_{-\infty}^{\infty}dq'\,{P\over q-q'}\,g(q')
   +{1\over 2}\,\langle f|g\rangle\,.
\]
In particular, for even $f$ and odd $g$, $\langle f|g\rangle$ vanishes, and
\[ \langle f|\kh'|g\rangle = -{i\over\pi} \int_0^{\infty}dq\,f^*(q)
   \int_0^{\infty}dq'\,\left({1\over q+q'}-{P\over q-q'}\right)g(q')
   \,,
\]
which gives eq.\rf{III.22} if $g$ coincides with $f$ on $(0,\infty)$. The
continuation from $\cal S$ to $L^2(-\infty,\infty)$ is performed as usual by
continuity, using the fact that $\cal S$ is a dense subspace in
$L^2(-\infty,\infty)$. 

\section{Study of the operator K}
\setcounter{equation}{0}

From the very definition of $K$ through the integral kernel \rf{III.24}, one
has
\[ (Kh)(q) = {1\over\pi}\int_0^\infty dq'\,{h(q')\over q+q'}\,.
\]
Let us put
\[ \hb(u) = \e{{\scriptstyle u}\over 2}h(\e{u})\,.
\]
Since $\int_{-\infty}^\infty du\,|\hb(u)|^2=\int_0^\infty dq\,|h(q)|^2$, the
correspondence $h\mapsto\hb$ defines a unitary mapping $L^2(0,\infty)
\rightarrow L^2(-\infty,\infty)$ and
\< \overline{\!Kh}\,(u) = \int_{-\infty}^\infty dv\,\,\Kb(u-v)\hb(v)\,,
   \label{B.1}
\>
where
\[ \Kb(u) = {1\over 2\pi\cosh{{\displaystyle u}\over 2}}\,.
\]
Then, another unitary mapping $L^2(-\infty,\infty) \rightarrow
L^2(-\infty,\infty)$, namely the Fourier transform
\[ \hbt(k) = {1\over\sqrt{2\pi}} \int_{-\infty}^\infty du\,\e{iku}\,
   \hb(u)\,,
\]
reduces the convolution product in \rf{B.1} to an ordinary product
\[ \widetilde{\overline{\!Kh}}(k) = \Kbt(k)\,\hbt(k)\,,
\]
where
\< \Kbt(k)\equiv\int_{-\infty}^\infty du\,\e{iku}\,\Kb(u) =
   {1\over\cosh\pi k}\,. \label{B.2}
\>
Therefore, the operator $K$ on $L^2(0,\infty)$ is unitarily equivalent to
the multiplicative operator $\rf{B.2}$ on $L^2(-\infty,\infty)$. The latter
is evidently a positive operator with purely continuous spectrum $[0,1]$. Its
generalized (non normalizable) ``eigenfunctions'' are 
\[ \hbt_s(k) = \delta(k-s) \qquad (s\in\R)\,,
\]
with ``eigenvalues'' $\lambda_s={1\over\displaystyle\cosh{\pi s}}$\,, and
their preimage in $L^2(0,\infty)$ are
\[ h_s(q) = {1\over\sqrt{2\pi q}} \e{-is\ln q}\,.
\]
Of particular interest for us is the extremal one, with ``eigenvalue''
$\lambda_0=1$
\[ h_0(q) = {1\over\sqrt{2\pi q}}\,.
\]
Of course, the corresponding maximal value $\gamma_0=1$ of $\gamma=\langle
h|K|h\rangle$ cannot be attained, but only approached arbitrarily close
through a family of normalizable functions mimicking ${1\over\sqrt{q}}$. For
instance, introducing two cutoffs, $\varepsilon$ at small $q$ and $L$ at
large $q$, and setting
\[ h_{\varepsilon,L}(q) = {1\over\sqrt{\ln {L\over\varepsilon}}}\,
 \chi_{(\varepsilon,L)}(q)\,{1\over\sqrt{q}}\qquad(\|h_{\varepsilon,L}\|=1)\,,
\]
one gets
\[ \langle h_{\varepsilon,L}|K|h_{\varepsilon,L}\rangle =
   1-{4\over\pi\ln{L\over\varepsilon}}\,\int_{\sqrt{\varepsilon/L}}^1dx\,
   {\arctan x\over x} =
   1-\mbox{O}\left({1\over\ln{L\over\varepsilon}}\right)\,,
\]
so that 
\renewcommand{\arraystretch}{.7}
$\build{\lim}{\begin{array}{cc} \scriptstyle\varepsilon\rightarrow0 \\
\scriptstyle L\rightarrow\infty \end{array}}
\langle h_{\varepsilon,L}|K|h_{\varepsilon,L}\rangle = 1$\,.
\renewcommand{\arraystretch}{1.4}

	Notice that one can keep $\varepsilon$ fixed (e.g.\ $\varepsilon=1$)
and let $L$ alone go to $\infty$ without changing anything (this is in fact a
consequence of the scale invariance of the operator $K$), or even choose a
family of less singular functions $h$, like
\[ h_L(q) = {1\over\sqrt{\ln(L+1)}}\,\theta(L-q)\,{1\over\sqrt{q+1}}\,.
\]

\end{document}